\documentclass[journal=nalefd,manuscript=letter]{achemso}
\setkeys{acs}{etalmode=truncate,maxauthors=10}

\usepackage{graphicx}
\usepackage{amsfonts}
\usepackage{amsmath}
\usepackage{natbib}
\usepackage{bm}
\usepackage{color}

\def\ai		{{\it ab initio}}

\def\ee         {electron--electron}
\def\ep         {electron--phonon}
\def\eg         {electron--photon}
\def\tdkr{TD--KERR}
\def\r{\rho}
\def\t{\tau}
\def\th{\theta}
\def\kk		{{\bf k}}
\def\w{\omega}

\renewcommand{\(}{\left(}
\renewcommand{\)}{\right)}
\def\ua{$\uparrow$}
\def\da{$\downarrow$}

\newcommand{\onlinecite}[1]{\hspace{-1 ex} \nocite{#1}\citenum{#1}}

\title{Ab initio calculations of ultra-short carrier dynamics in
2D materials: valley depolarization in single-layer WSe$_2$}

\author{Alejandro Molina-S\'{a}nchez}
\affiliation{Physics and Materials Science Research Unit, University of
Luxembourg, 162a avenue de la Fa\"iencerie, L-1511 Luxembourg, Luxembourg}
\email{alejandro.molina@uni.lu}

\author{Davide Sangalli}
\affiliation{CNR-ISM, Division of Ultrafast Processes in Materials (FLASHit),
Area della Ricerca di Roma 1, Monterotondo Scalo, Italy}

\author{Ludger Wirtz}
\affiliation{Physics and Materials Science Research Unit, University of
Luxembourg, 162a avenue de la Fa\"iencerie, L-1511 Luxembourg, Luxembourg}

\author{Andrea Marini}
\affiliation{CNR-ISM, Division of Ultrafast Processes in Materials (FLASHit),
Area della Ricerca di Roma 1, Monterotondo Scalo, Italy}

\date{today}


\begin{document}

\begin{abstract}
In single-layer WSe$_2$, a paradigmatic semiconducting transition metal dichalcogenide, 
a circularly polarized laser field can selectively excite electronic transitions
in one of the inequivalent $K^{\pm}$ valleys. Such selective valley population
corresponds to a pseudospin polarization. This can be used as a degree of freedom in  a ``valleytronic'' device
provided that the time scale for its depolarization is sufficiently large.
Yet, the mechanism behind the valley depolarization still remains heavily debated.
Recent time--dependent Kerr experiments have provided an accurate way to
visualize the valley dynamics by measuring the rotation of a linearly polarized probe pulse applied after a circularly polarized pump pulse.  We
present here a clear, accurate and parameter--free description of the valley dynamics. By using an atomistic, 
\ai\, approach we fully disclose the elemental mechanisms that dictate the depolarization effects.
Our results are in excellent agreement with recent time--dependent Kerr
experiments. We explain the Kerr dynamics and its temperature dependence in terms of
electron--phonon mediated processes that induce spin--flip inter--valley transitions. 
\end{abstract}

\maketitle

The semiconducting transition metal dichalcogenides\,(TMDs) are layered
materials with a two--dimensional honeycomb crystal structure and a peculiar
electronic structure: in the single-layer, the direct
band gap located at the corners of the hexagonal Brillouin zone gives rise to
two inequivalent {\it valleys}\cite{Cao2012,Mak2012,Zeng2012}, $K^{+}$ and $K^{-}$.  An
electron/hole being able to move between these two valleys can be described as a
two--component pseudo--spinor.  Therefore, the valley degree of freedom is also
known as pseudospin,  a new quantum number that distinguishes between carriers
in different valleys\cite{Xu2014}.  The pseudospin is able to carry information and, also,
to generate a valley Hall effect\cite{Urbaszek2015,Lee2016}, turning TMDs into promising building blocks of the new emerging field of valleytronics and
potential valleytronic--based devices.\cite{Schaibley2016} 

The generation and persistence\cite{Glazov2014,Yang2015,Rivera2016,Ye2016} of a carrier imbalance 
between $K^{\pm}$ valleys is the key requirement for the efficient use of the
pseudospin\cite{Rycerz2007,Xiao2012,Xu2014}. Ultrafast spectroscopy is an advanced and
accurate experimental technique able to investigate the pseudospin real--time
dynamics\cite{Marie2015}, identifying the key mechanisms that drive
the valley depolarization. Time-dependent Kerr rotation spectroscopy (TD-KERR) measures the 
Kerr angle, which is directly proportional to the valley polarization.\cite{Poellmann2015,Nie2014,Zhukov2007} The TD-KERR 
experiments on TMDs\cite{Zhu2014,DalConte2015} show very different decay times
of the valley polarization, spanning from pico to nanoseconds. The large
difference has been tentatively attributed to intrinsic doping or localized
defect states.\cite{Yang2015} 

While both theoretical models and first--principles simulations have described the key
physical properties of TMDs and how valley polarization is induced by circularly polarized
pulses, the microscopic understanding of the subsequent valley depolarization is still strongly debated,
despite the numerous experimental data
available.~\cite{Winnerl2011,Mai2014a,yu_valley_2014,Glazov2015,Sie2015,Hao2016}

One of the most renowned mechanism proposed in the literature to explain the valley depolarization is the 
electron-hole exchange\,(eh-X) mechanism\cite{yu_valley_2014}.
In the eh-X case the circularly polarized pump pulse is assumed to excite an independent electron--hole pair in a single valley.
This pair is subsequently scattered to the other valley via virtual transitions caused by 
the bare electron--hole exchange interaction. Temperature is introduced by assuming that the initial pairs are distributed according to 
a Boltzmann distribution at a given temperature~\cite{Hao2016}. In the eh-X mechanism the out--of--equilibrium dynamics of the
 material is reduced to the solution of the equation of motion for  a 2$\times$2 matrix and the scattering part of the dynamics
 is introduced using adjustable, {\em ad--hoc} lifetimes.\cite{Maialle1993}

Although appealing, the eh-X approach is based on a crucial 
approximation. The initial non-equilibrium state, which should mimic the state
after the action of the pump pulse, is assumed to be uncorrelated. However,
it is well known that the optical properties of TMDs are characterized by a
strong electron-hole interaction\cite{Molina-Sanchez2015,Molina-Sanchez2016}. This is easily visualized in the 
optical spectrum of  WSe$_2$, dominated by excitonic peaks that are 
composed by several, correlated and coherent electron--hole
pairs~\cite{Cudazzo2011,Latini2015}. Thus the electron--hole exchange interaction
should not be considered on the uncorrelated state, but on the correlated state
created by the pump. To this end a correct modeling should include the pump
pulse in the simulation.

Such a coherent approach, where excitonic effects are included from the beginning together with realistic scattering mechanisms, does exist.
It is based on an atomistic description of the material and it is rooted on the well--know merging of Many--Body techniques with
Density--Functional--Theory~\cite{Onida2002} (also known as \ai\, approaches).
An \ai\, non--equilibrium dynamics is not based on a simplified model and on adjustable parameters. All ingredients of the method are calculated
from the very elementary atomic structure of the material. This make \ai\, approaches predictive.
Therefore, an \ai\, simulation can provide a quantitative result and thus a convincing
answer to the mechanism which drives the valley depolarization in TMDs.

Such a non--equilibrium \ai\, simulation must, however, include several key ingredients to be accurate and convincing.
First of all, both the pump pulse and the scattering mechanism need
to be included in a first--principles manner and going beyond simple models based on parameters extracted {\em a posteriori} from the experimental results.
Second, electronic correlation needs to be included to account
for the strong electron--hole interaction during the absorption of the pump pulse.
Finally, the spin--orbit interaction also needs to be fully described to couple the spin and pseudospin
polarization dynamics~\cite{Xiao2012,Schaibley2016}.
In this work we use such a set of theoretical and numerical tools to model
the valley depolarization in two--dimensional TMDs in a clear and convincing way. 

\begin{figure*}
\includegraphics[width=14 cm,scale=0.3]{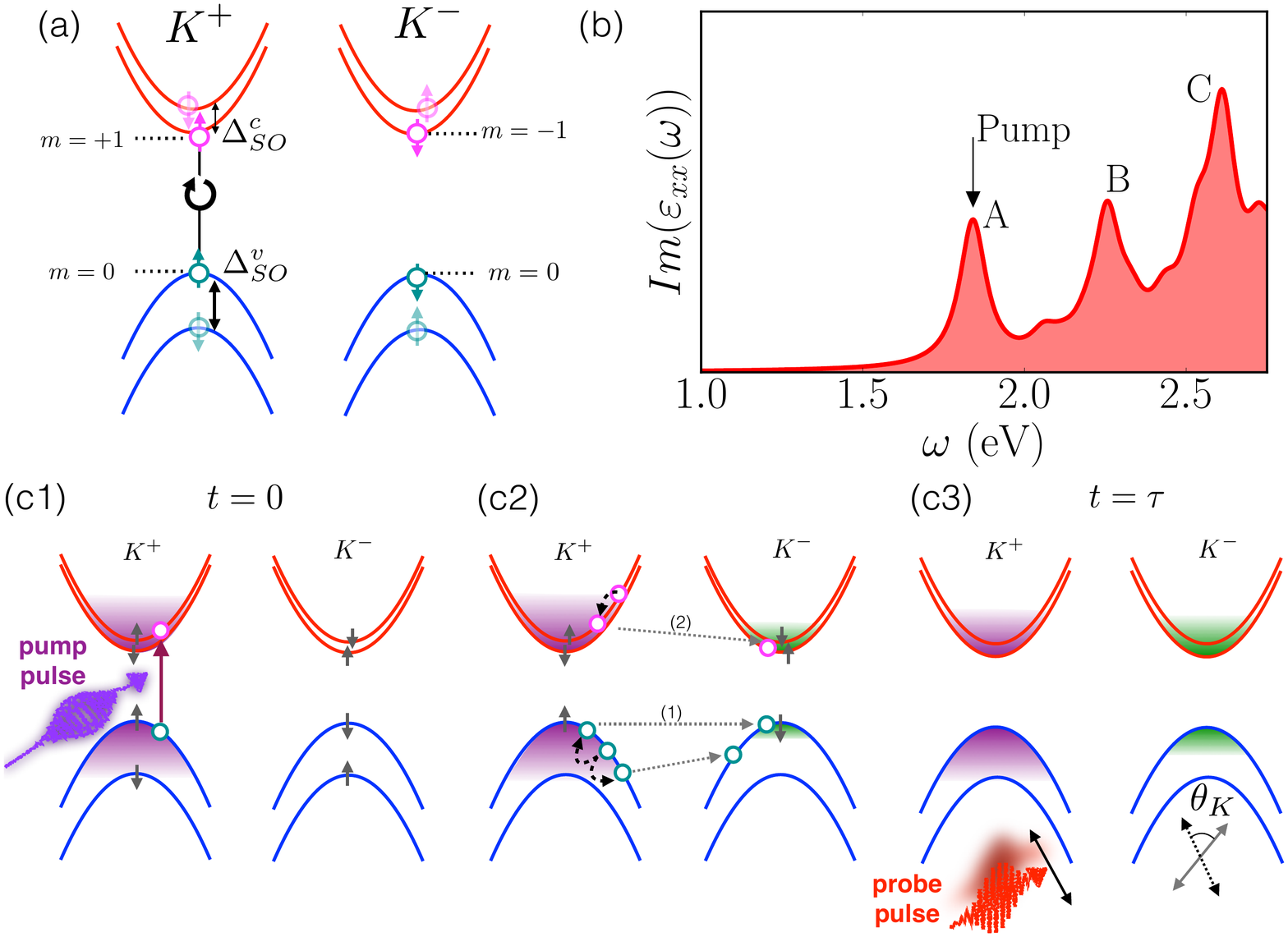}
\caption{The different steps involved in a \tdkr\, experiment in a single--layer
TMD. (a) Schematic representation of the selection rules for circularly
polarized light at the 
$K^{\pm}$ valleys. (b) Optical absorption spectrum. (c1--3) Steps of the experiment.  (c1) A circularly polarized pump promotes electrons in the conduction
bands only in one valley, creating a finite pseudospin polarization. The pump
frequency is tuned to the A excitonic absorption line. (c2) The carriers
scatter from one valley to the other in order to equalize the population of the
$K^{\pm}$ valleys, whose density is represented by violet ($K^+$) and green
($K^-$). (c3) After a delay $\tau$ a linearly polarized laser is used to probe the valley 
polarization dynamics by measuring its polarization axis rotation. This rotation is caused by the Kerr effect, induced by the magnetization associated with the
pseudospin polarization. The exponential decay
defines the valley polarization lifetime that plays a crucial role in the
valleytronic properties of the material. In the frames\,(c1--c3) we represent the elemental scatterings involved in the time--dependent Kerr dynamics. In
general, electrons and holes follow different paths depending on the specific properties of the electronic structure.}
\label{fig-intro}
\end{figure*}

\begin{figure*}
\includegraphics[width=14 cm]{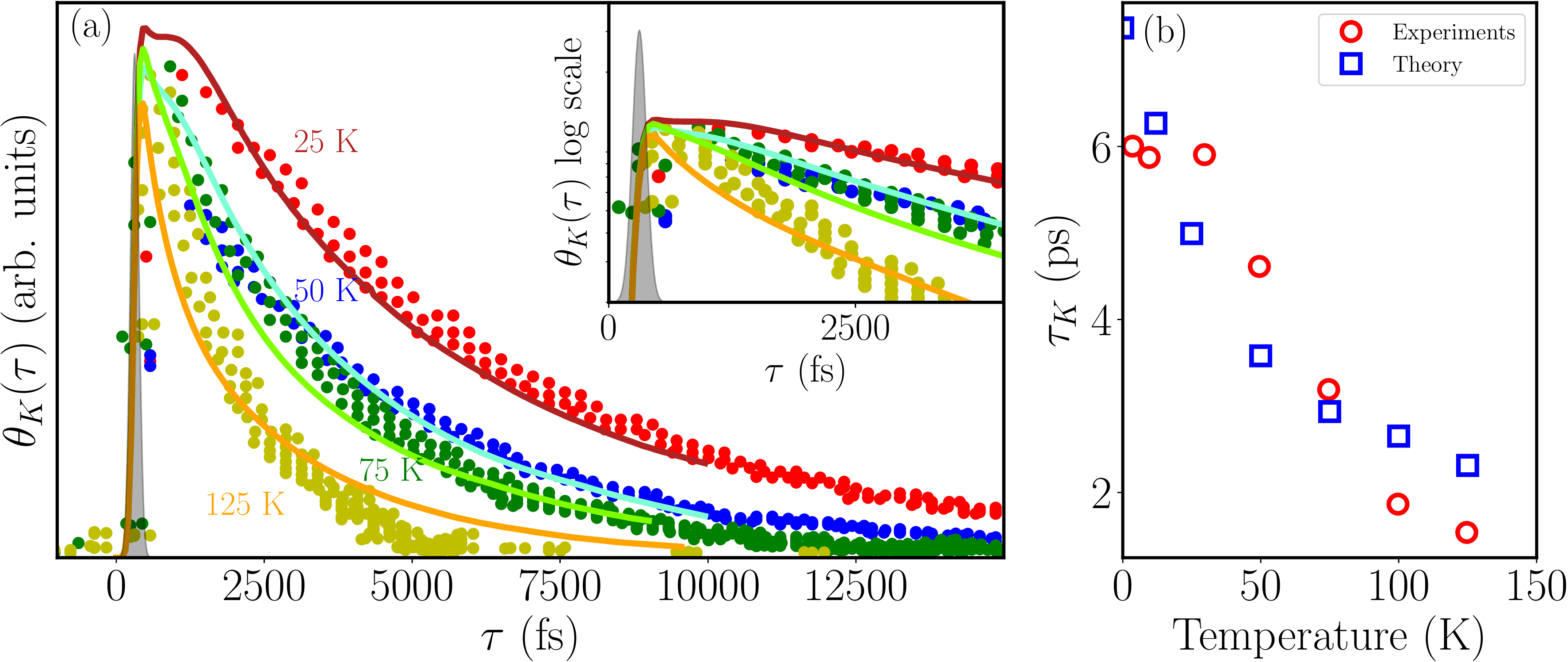}
\caption{(a) Kerr angle $\th_K(\t)$ for several temperatures. The
theory\,(lines) is compared with experiments\cite{Zhu2014}\,(dots). The shadow gray region represents the
pumpe pulse. In the inset the same comparison is presented on a logarithmic scale.
The theoretical Kerr lifetime $\t_K$, obtained from a fit with a single exponential, is compared with the
experimental values in the frame (b).}
\label{fig-kerr}
\end{figure*}

We consider a \tdkr\, experiment performed on a single layer WSe$_2$ (see Fig.
\ref{fig-intro} and \ref{fig-kerr}). The Kerr angle shows a clear
exponential decay, with a lifetime that increases as the temperature drops and reaches a low--temperature plateau.
We use an \ai\, implementation of the Kadanoff--Baym out--of--equilibrium equations to describe the 
dynamics in a Kohn--Sham basis. Our theoretical simulations describe the creation of a valley polarization and perfectly reproduce 
its experimental decay, which allows us to give a complete and exhaustive explanation of the physics involved. We find that the
valley depolarization is mainly caused by spin--flip transitions mediated by
\ep\, interaction. We also successfully explain the temperature dependence of the Kerr angle.

\textbf{An atomistic approach to the Time--dependent Kerr effect}. In a prototype TD-KERR experiment (Fig.~\ref{fig-intro}.(c1-c3)), a sharp circularly polarized
laser pulse is used to pump the material by photo--exciting electrons to the
empty conduction states\,(step c1), creating the carrier imbalance at the
inequivalent valleys\cite{Mak2012,Zeng2012,Wu2013,Xie2016}.
Circularly polarized light mainly excites electron--hole pairs in a given
valley, as summarized in Fig.~\ref{fig-intro}(a). While selection rules impose
spin conservation, the azimuthal quantum number ($m$) changes from $m=0$ in valence
to $m=+1$ ($m=-1$) in conduction at $K^+$ ($K^-$) when left 
(right) circularly polarized light is used, respectively\cite{Cao2012}. The imbalance between the valley population
entails a slight magnetization of the layer, responsible for the Kerr angle rotation. After a time delay
$\tau$, the system is probed with a second weak field, linearly polarized. If 
there is a finite valley polarization, i. e., the carrier imbalance persists, 
the probe axis will be rotated by an angle $\theta_K (\t)$ (steps c3).

The TD-KERR experiment on WSe$_2$ single layer\cite{Zhu2014} (see Fig.
\ref{fig-kerr}) clearly shows that the Kerr angle decays exponentially, with a
dependence on the delay that can be well described by a lifetime of the order of
a few pico--seconds (ps). The lifetime decreases with
temperature. We aim to find the cause of this decay and how to describe
the temperature dependence.

Our approach is based on a fully \ai\, framework that merges the
non--equilibrium Green's function\,(NEGF) theory with the well--known Density
Functional Theory\,(DFT) (See Methods in Supporting Information). DFT 
is used to calculate all ingredients of the NEGF dynamics: the electronic and phononic states, the \ep, \ee\, and \eg\, scattering amplitudes. 

As it will become clear in the following, the speed of the valley depolarization is
strongly influenced by the spin--orbit coupling\,(SOC) of the electronic
bands near $K$ (see Fig. \ref{fig-intro} (a)). In particular, the SOC of the valence band maximum 
is very strong ($\Delta^{v}_{SO}=0.49$ eV) and leads to opposite spin polarization of the VB maxima in the $K^{\pm}$ valleys. This slows down considerably the electron-phonon mediated 
scattering of holes between the valleys. The
fact that $\Delta^{c}_{SO}\ll\Delta^{v}_{SO}$ means that the SOC
impact on the electron dynamics is much less pronounced. In any case, a careful
simulation of \tdkr\, dynamics requires a precise calculation of
the band structure using full spinorial wave functions. 

The time--dependent dynamics is obtained by solving the Kadanoff--Baym equation\,(KBE) for the time evolution of the single--time density--matrix,
$\r_l(\t)$. $\r_l$ is written in the Kohn--Sham 
basis for the electron--hole pairs, $l=(nm\kk)$. Here $n$ (and $m$) 
are spinorial band indices and $\kk$ is the
crystal momentum. The KBE\cite{Marini2013,Sangalli2015,Sangalli2016,Melo2016} reads
\begin{align}
\partial_{t} \r(\t)_l =  \left. \partial_{t} \r(\t)_l\right|_{coh} +  \left. \partial_{t} \r(\t)_l\right|_{coll},
\label{eq-motion}
\end{align}
where $\left. \partial_{t} \r_l\right|_{coh/coll}$ are the coherent/collision
terms of the dynamics, respectively.
 
\textbf{The interaction with the pump pulse}. The term $\left. \partial_{t} \r_l\right|_{coh}$ describes 
the interaction of the pump laser with the system and embodies correlation effects 
induced by the
\ee\, interaction. It is approximated within the statically screened exchange
approximation~\cite{Attaccalite2011} that describes correctly 
excitonic effects coherently with the well--known and widely used
Bethe--Salpeter equation~\cite{Onida2002}. The present form of $\left. \partial_{t}
\r_l\right|_{coh}$ has been shown to be reliable in several works about 
out--of--equilibrium dynamics~\cite{Pogna2016,Sangalli2016,Perfetto2016} and non--linear optical
properties~\cite{PhysRevB.89.081102,PhysRevB.88.235113,C5CP00601E}.

The first step in the simulation of the TD-KERR spectra is to calculate the optical
absorption. Neglecting the collision term, from the solution of the KBE, the frequency
dependent dielectric function $\varepsilon^{EQ}_{xx}(\w)$ 
can be extracted by Fourier transform (Fig. \ref{fig-intro}(b)). This corresponds to 
a calculation of the absorption spectrum with the static Bethe--Salpeter equation~\cite{Attaccalite2011}.
The spectrum shows a clear excitonic peak $A$, and a secondary peak $B$.
The $A-B$ splitting is induced by the large SOC splitting of the valence bands
of single--layer WSe$_2$ \cite{Molina-Sanchez2015},
$\Delta_{SO}^v$ in  Fig. \ref{fig-intro}(a). The good description of the optical
properties ensures that the
pump pulse is correctly absorbed.

The experimental \tdkr\, spectra we want to reproduce\cite{Zhu2014} are measured with a circularly polarized laser pump
pulse\cite{Mak2012} with  100\,fs full width at half maximum of the envelope profile.
The pump intensity is $10^5$ kW$/$m$^2$ and it is energetically centered on the $A$ exciton. The experimental
intensity corresponds to a maximum carrier density of 10$^{-12}$ cm$^{-2}$ and allows us to work in the low--intensity
regime, formally defined in Refs. \onlinecite{Sangalli2016,Melo2016}.

\textbf{Valley polarization dynamics: \ep--mediated spin--flips.}
The driving force of the dynamics is carried by the collision term $\left. \partial_{t}
\r_l\right|_{coll}$. It describes dephasing and scattering 
induced by \ep\, (and possibly \ee\, and \eg\,) processes,
and it permits a clear estimation of their role and interaction with the spin
and pseudospin polarizations. Temperature in our work enters naturally via
the Bose distribution of the phonons. It affects the interaction with the pump,
as discussed in Ref. \onlinecite{Marini2008}, and the subsequent dynamics, as
established in Refs. \onlinecite{Marini2013} and \onlinecite{Sangalli2015a}.
Equation \eqref{eq-motion} has been 
derived from the very general KBE by using a series 
of approximations that we have summarized in the Methods section of the
Supporting Information and that has
been extensively discussed in Refs.\,\onlinecite{Marini2013,Melo2016}. The
collision term has been used to describe
inter--valley scattering in Silicon~\cite{Sangalli2015} and the complex interplay between
dissipation and correlation in the ultra--fast formation of Fermi
distributions~\cite{Sangalli2015} of the photo--excited carriers.

The time dependent density matrix, solution of Eq.~\eqref{eq-motion}, provides
the non--equilibrium (NEQ) occupations of the electronic levels, $f_{n\kk}$ as a function of the delay time $\tau$
\begin{equation}
f_{n\kk}(\t) = \r_{nn\kk}(\t).
\label{eq-occ}
\end{equation}
From the NEQ occupations we can calculate several time--dependent observables.
Our strategy relies on the idea that the physics we are
interested in occurs in a time window that does not overlap with the pump laser pulse.
Therefore we are in a quasi--adiabatic regime\cite{Perfetto2015},
where the density--matrix off--diagonal matrix elements have been dephased and are zero. All physical observables can
be made time--dependent by expressing them in terms of the time--dependent
occupation functions. This approach has been successfully applied to the dielectric
tensor $\varepsilon_{\alpha\beta}(\omega,\tau)$\cite{Sangalli2016} (with $\alpha,\beta=\{x,y\}$)
and used to describe the transient absorption experiments in MoS$_2$\cite{Pogna2016} and bulk Silicon\cite{Sangalli2016}. 

\textbf{The time and temperature dependent  Kerr signal.}
In the TD-KERR spectra the typical order of magnitude of the Kerr angle is milliradians~\cite{DalConte2015}.
In this small-angle limit, 
the Kerr angle is directly related to the non-diagonal matrix elements of  the dielectric function by the equation~\cite{Sangalli2012}
\begin{equation}
  \th_K(\w,\t)=\Re[ \frac{-\varepsilon_{xy}(\w,\t)}{(\varepsilon_{xx}(\w,\t)-1)\sqrt{\varepsilon_{xx}(\w,\t)}} ],
\label{eq-kerr}
\end{equation}
with $\w$ the probe frequency. The spin in the Kerr angle enters via
the SOC coupling between spin and the spatial part of the wave
function\cite{Sangalli2012}. The dielectric function is expressed as
\begin{equation}
    \varepsilon_{\alpha\beta}(\w,\t)=\delta_{\alpha\beta}-                          
    \frac{4\pi e^2}{\hbar}\sum_{l\,l'}  
        (x^\alpha_{l})^*\,
       \frac{f_l(\t)} {\hbar\omega-\epsilon_l+i\eta}\,
         x^\beta_{l'}
\end{equation}
with $x^\alpha_l=x^\alpha_{cv\mathbf{k}}$ the optical dipole matrix elements, $f_l(\t)=f_{v\mathbf{k}}(\t)-f_{c\mathbf{k}}(\t)$ the difference
between the electronic occupations and $\epsilon_l=\epsilon_{c\mathbf{k}}-\epsilon_{v\mathbf{k}}$ the difference
between the electronic energies.
The off-diagonal term, i.e. $\alpha\neq\beta$ is non-zero only if there is a breaking of symmetry which,
in the  present case, is induced by the circularly polarized pump pulse. It is worth to
notice that the dielectric function is here defined ``per surface unit'', with
the surface factors absorbed in the definition of the dipoles.
Our
approach fully accounts for the contributions of SOC and the difference in valley
occupation contribution to the Kerr angle. The calculated $\th_K(\t)\equiv \th_K(\w=E_{max},\t)$ is compared with the
experiment results of Ref.~\onlinecite{Zhu2014} in Fig.~\ref{fig-kerr}. As in the experiment we tune
the probe frequency to $E_{max}$, i.e. the maximum of $\varepsilon_{xy}(\w,\t\simeq 0)$, just after the pump pulse.
The decay of the simulated Kerr signal is simultaneous with the end of the pump
pulse (see Supporting Information for details). We obtain a good agreement if we
fit the theoretical Kerr curves with a single exponential, as represented in
Fig. \ref{fig-kerr}(b). The predicted $\th_K\(\t\)$ nicely agrees with the experimental results
in the long time regime for all the measured temperatures, $T=25-120$\,K. We can
also see that both theory and experiment exhibit a finite value  when $T\rightarrow 0$ K. This finite value is due to the
atomic zero--point vibrations. Only  in the high temperature regime we observe a deviation for $\t>4$ ps.

\begin{figure}
\includegraphics[width=7 cm,scale=0.4]{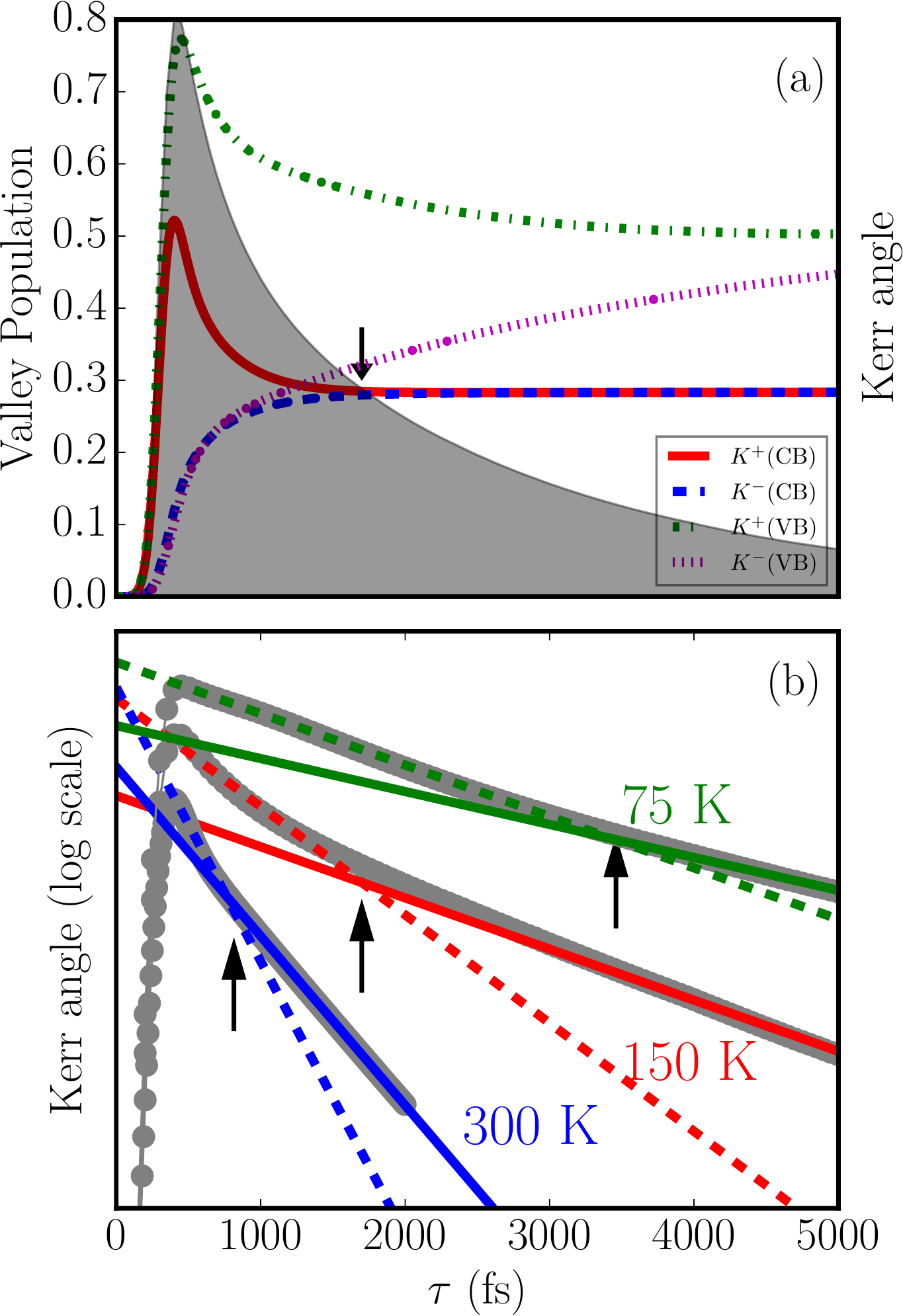}
\caption{(a) $K^{\pm}$ valleys population splitted in valence\,(VB) and conduction\,(CB) contributions at $T=150$\,K. The corresponding Kerr angle is
represented by the shadow region. From the very different raise of the $K^{-}$
population the different dynamics of electrons and holes appear clearly. For 
 $\t\approx$\,1\,ps is clear that, while the CB dynamics is already saturated, the VB holes are still migrating from one valley to the
other. (b) $\theta_K\(\t\)$ for temperature $T=75,150,300$\,K in a logarithmic scale. We have fitted the Kerr angle decay with two straight lines in order to
represent the fast ($\t^{fast}_K$) and  slow ($\t^{slow}_K$) components. The
arrows indicate the transition from fast to slow decay, which corresponds
exactly to the delay where the two CB components of the $K^{\pm}$ valleys are equally populated.}
\label{fig-kerr-dyn}
\end{figure}

As schematically represented in Fig. \ref{fig-intro}(c2), the decay of the Kerr
angle and of the valley polarization depends on the scattering rates in the 
valence\,(VB)
and conduction bands\,(CB). The large difference in the SOC
splitting for VB and CB suggests different scattering rates for electrons and
holes, which can be visualized as two characteristic lifetimes: one fast, $\tau^{fast}_K(T)$
and the other slow, $\tau^{slow}_K(T)$. Fig. \ref{fig-kerr-dyn}(a) show the VB and
CB valley populations at the inequivalent $K^{\pm}$ points for $T=150$ K. 
The carrier inter--valley scattering is much faster for the CB, compared to
the VB. While at 150 K and $\t=1$\,ps the $K^{\pm}$ CB populations are
the same and the CB contribution to the pseudospin polarization vanishes, the VB carriers are still
slowly moving out of the $K^+$ valley. We can see a clear signature of this different dynamics in the logarithmic representation of the Kerr signal,
Fig. \ref{fig-kerr-dyn}(b). The values of the fast and slow lifetimes are shown
in Table \ref{table-lifetimes}. Indeed, the transition from fast to slow decay
corresponds exactly to the point where the CB  $K^{\pm}$ populations are the
same and the entire Kerr dynamics is dictated by the slower VB contribution. As a consequence the decay lifetime changes from $\t^{fast}_K\(T\)$ to
$\t^{slow}_K\(T\)$. In agreement with the experimental results, with increasing
temperature, the transition from fast to slow decay occurs at earlier times.
Notice that in the model
of Ref. \onlinecite{Zhu2014} the lifetimes diverge at low temperature, while our
calculations capture the finite zero--temperature lifetime. Our approach gives also the full dynamics
of the Kerr angle as function of time, in which we observe two different
lifetimes, a result which a model based in a single lifetime cannot reproduce.
The experimental data shows a very scattered distribution which does not allow a
clear identification of the faster dynamics.

\begin{table}
\center
\label{table:sl}
\begin{tabular}{lccccc}
\hline
T (K)  & $\t^{fast}_K$ (ps) & $\t^{slow}_K$ (ps) \\
\hline
75     & 2.59               & 3.73               \\
150    & 0.89               & 2.50               \\
300    & 0.37               & 0.75               \\
\hline
\hline
\end{tabular}
\caption{Fast and slow lifetimes ($\t^{fast}_K$ and $\t^{slow}_K$) for
temperatures of Fig. \ref{fig-kerr-dyn}.}
\label{table-lifetimes}
\end{table}

\textbf{Relaxation paths in the hexagonal Brillouin zone.}
While the \ep\, and \ee\, channels can contribute to the inter--valley scattering by using finite momentum phonons 
and electron--hole pairs (as sketched in Fig. \ref{fig-intro}(c2)), the \eg\,
interaction cannot. Since photon--assisted transitions are vertical in
momentum space, photon--assisted inter--valley scattering is unlikely. Experimental data in time-resolved photoluminescence seems to indicate 
radiative lifetime that increase with
temperature.\cite{Zhang2015,Robert2016} The same trend is predicted in
\ai\, simulations~\cite{Palummo2015}. This is in contrast to the trend observed
in the \tdkr\, experiment. We thus conclude that electron-photon interaction is
not relevant in the present context.

We have performed simulations including both \ee\, and \ep\, interactions. We found the \ep\, channel to be dominant with the \ee\, scattering only giving a
minor contribution.  The reason is that the \ep\, and \ee\, channels have a different dependence on the photo--excited carrier density and energy.  Indeed, the \ee\,
channel can be split into two terms. The first, or equilibrium term, is activated only when carriers are excited with sufficient energy to generate cascade Auger
processes internal to the low--energy valence/conduction bands~\cite{Marini2013,Bernardi2014}. This is not the case of the present \tdkr\, experiment since (bound)
electrons and holes are created near the conduction band minimum and valence band maximum. The second term, or non--equilibrium term, does not have any
energy constraint but scales either quadratically or with the third power of the
carrier density. For increasing number of carriers the
electron-electron channel is expected to gain importance. Verification of the role of
electron-electron interaction can be done by increasing the pump intensity. In 
contrast the \ep\, channel is always active and scales
linearly with the density. It is, therefore, dominant in the experimental range of carrier densities (10$^{-12}$\,cm$^{-2}$, in Ref.\cite{Zhu2014})
studied here.

\begin{figure*}
\includegraphics[width=13 cm]{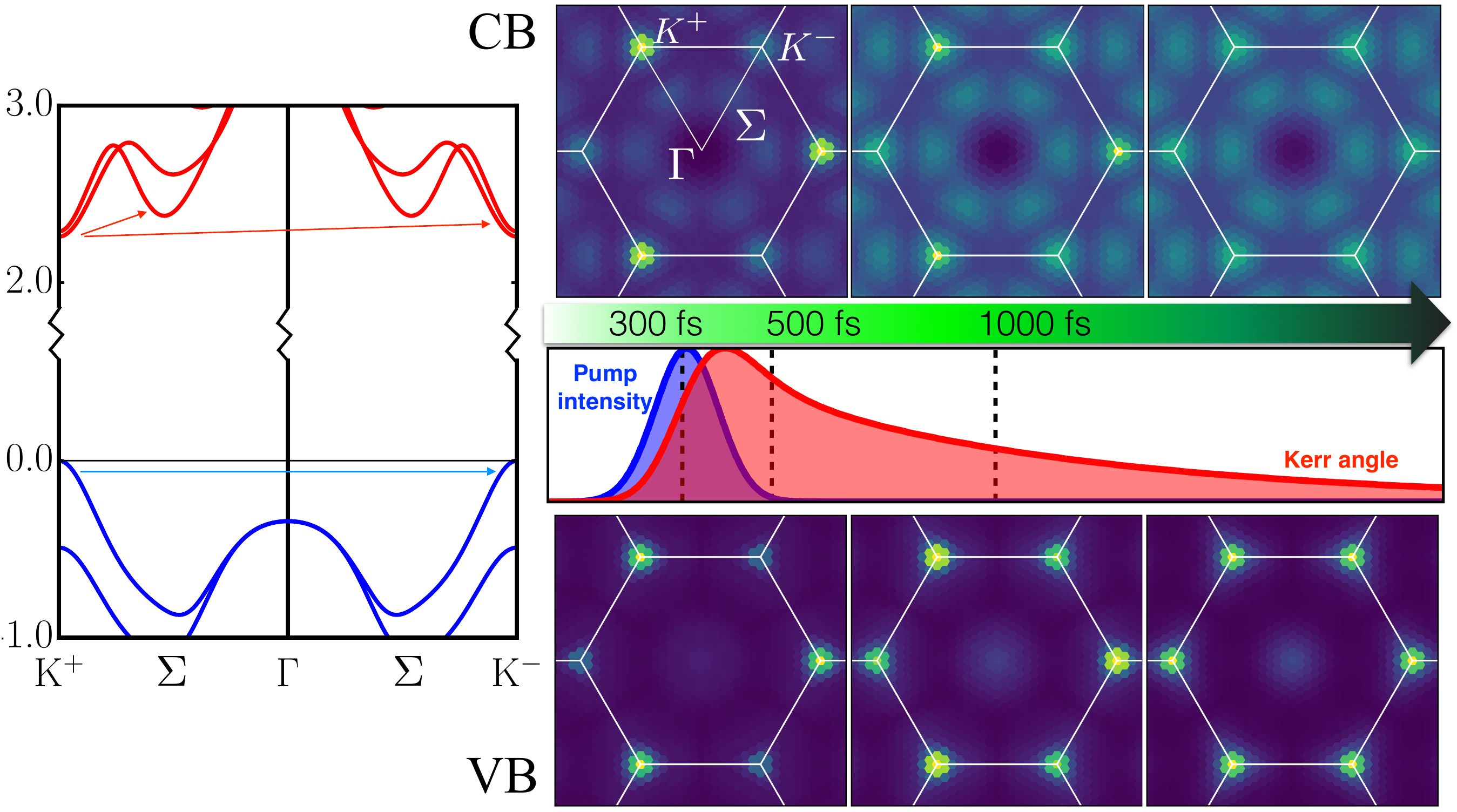}
\caption{Band structure of WSe$_2$ (left panel), with 
the relevant scattering paths labelled by arrows. In the right frame we show the
occupations of all photo--excited
electrons (CB) and holes (VB), projected on the hexagonal plane, as
snapshots corresponding to $\t=300,500$\,fs and $\t=1$\,ps, at $T=300$\,K. The very different dynamics of VB and CB 
is evident. In both cases the final equilibrium is
reached by spin--flip and pseudospin--flip transitions but using different
kinematic routes (a full movie can be found in the Supporting Information).}
\label{fig-density}
\end{figure*}

The contrasting CB and VB dynamics can now be fully understood thanks to our
microscopic approach. Figure \ref{fig-density} shows the electronic density of the CB
and VB as two--dimensional snapshots in $\bf k$-space corresponding to
$\t=300,500$\,fs and $\t=1$\,ps. The temperature is $T=300$\,K. We
have labelled the K$^{\pm}$ valleys and the $\Sigma$ high symmetry line. The
band structure along $K^+ -- \Gamma -- K^-$ is shown in the left frame.

We clearly see that the circularly polarized pump initially populates only the K$^+$ valley thus creating a pseudospin (and spin) polarized excited state.  As
the delay $\t$ increases, also the K$^-$ valley acquires a finite population but following a different dynamics for the valence and conduction bands. Electrons
and holes can jump from one valley to the other following two possible paths:
either by (i) direct spin--flip
\cite{spinflip} or by (ii) conserving spin and overcoming
the SOC splitting
which exists between the \ua\,(\da) state at $K^+$ and the \ua\,(\da) state at
$K^-$ (see Fig. \ref{fig-intro} step c2). 
The conduction band SOC splitting is small ($\Delta^c_{SO}=30$\,meV, see
Fig. \ref{fig-intro}) and electrons can easily jump quickly to the K$^-$,
emitting phonons without spin--flip in a first instance and then equilibrate the
population of the two conduction bands.

On the other hand, the valence band holes have a slower dynamics. The SOC splitting for the VB states at $K^{\pm}$ is $0.491$\,eV, a barrier which cannot be
overcome via the emission of phonons.  The only option for a hole at $K^+$ to scatter directly into $K^-$ is to flip its spin.  A direct process where the spin
is flipped by the emission of a phonon is possible in WSe$_2$ because of the spinorial nature of the electronic states. In practice this means that every electron
(and hole) has both an $\uparrow$ and $\downarrow$ spin component and a single electronic transition is from a spinor to another and does not, in general,
preserve the spin.  However such a direct process has a very small transition rate (see SI for the representation of the electron-phonon matrix elements).
Therefore, holes follow a very slow dynamics in comparison with electrons even
when the temperature is as high as $T=300$\,K. We recall here the importance of
including full spinorial wave functions. Otherwise the scattering rate of the VB
would be similarly fast as the one of the CB. Our simulations show that the
$\Gamma$ point as an intermediate step in the relaxation plays only a very minor
role.\cite{Mai2014a}. The relative energy between $K$ and $\Gamma$ is 0.35 eV, which makes unlikely
such scattering path at experimental temperatures. 
Acoustic and optical phonon branches contribute with similar weight to the
scattering. From Ref. \cite{Molina-Sanchez2016} we learned that conduction and
valence states around the $K$ point couple to both acoustic and optical phonons,
consistent with the current results.

The picture provided by the present simulations rules out any other mechanisms
based on solely electronic degrees of freedom. We already pointed out, in the introduction, why the
electron--hole exchange interaction mechanism~\cite{yu_valley_2014,Maialle1993} alone
cannot explain the valley depolarization. The electron--hole exchange interaction is fully included in the 
 $\left. \partial_{t} \r_l\right|_{coh}$ term of Eq.\eqref{eq-motion}. But only the scattering term is responsible
for the valley depolarization, with the coherent part only describing the correct coupling with the 
laser pulse.
The present simulations, instead, points to a dominant role played by the
electron--phonon scattering\cite{Kioseoglou2016}.

\textbf{Conclusions} Our \ai\, simulations of the \tdkr\, angle accurately reproduces the experimental results for all  temperatures and time ranges.  The use of an atomistic approach
allows us to explain the mechanisms that drive the dynamics. We demonstrate that
the \ep\, interaction is the driving force of the valley depolarization and
motivate the features observed experimentally with a microscopic analysis of the scattering paths of the carriers.

The two key experimental facts are reproduced and explained: (i) electrons and holes tend to equalize the population in the $K^{\pm}$ valleys, and (ii) the
valley relaxation lifetime drops as the temperature increases. 

The photo--excited electrons and holes follow very different dynamics to reach
the final equilibrium configuration. Due to the small
spin--orbit splitting of the conduction bands near the $K^{\pm}$ valley minima and to the presence of several local minima in the Brillouin zone the
photo--excited electrons scattering is enhanced and the carriers are quickly
spread over wide regions of the Brillouin zone.  On the contrary, the large valence spin--orbit
splitting hinders efficient scattering processes. The holes $K^+\rightarrow\,K^-$ transition is bound to be a slow cascade process, involving mainly
direct valley to valley transitions.

The calculated ultra-short valley depolarization times correspond to a comparison
with intrinsic or low-doped single-layer WSe$_2$. Similar lifetimes have been observed in
intrinsic samples of single-layer Mo$_2$\cite{DalConte2015}. Differing lifetimes
that are attributed to doped samples will be the topic of future
work.

The key feature of the present \ai\, simulation scheme is the use of a fully spinorial basis.  This is a
mandatory feature to gain a successful description of the pseudospin dynamics: all ingredients (band structure, vibrational modes, \ep\, matrix elements) must be
calculated including spin--orbit effects. This rules out any use of simplified, few--bands models. A precise, predictive and accurate description
requires an atomistic treatment.

The present scheme provides a unique, parameter--free and
universal approach to interpret and predict several time--dependent experiments, like time--dependent Kerr, time-dependent photo--reflectance and, in general,
any pump-probe technique in the scale of femtoseconds to picoseconds.

\section*{Acknowledgements}
A. M.-S.\ and L.W. acknowledge support by the National Research Fund, Luxembourg (Projects C14/MS/773152/FAST-2DMAT and INTER/ANR/13/20/NANOTMD). AM
acknowledges the funding received from the European Union project MaX {\it Materials design at the eXascale} H2020-EINFRA-2015-1, Grant agreement n. 676598 and
{\it Nanoscience Foundries and Fine Analysis - Europe} H2020-INFRAIA-2014-2015, Grant agreement n. 654360.
The simulations were done using the HPC facilities of the University of Luxembourg~\cite{VBCG_HPCS14}. The authors declare no competing financial interests.

\begin{suppinfo}
The Supporting Information contains additional details about the rate equations
implemented for the carrier dynamics simulation, the electronic band structure of single-layer WSe$_2$, the time-dependent Kerr
spectra and the electron-phonon matrix elements.
\end{suppinfo}

%\bibliography{biblio}{}

\begin{mcitethebibliography}{55}
\providecommand*\natexlab[1]{#1}
\providecommand*\mciteSetBstSublistMode[1]{}
\providecommand*\mciteSetBstMaxWidthForm[2]{}
\providecommand*\mciteBstWouldAddEndPuncttrue
  {\def\EndOfBibitem{\unskip.}}
\providecommand*\mciteBstWouldAddEndPunctfalse
  {\let\EndOfBibitem\relax}
\providecommand*\mciteSetBstMidEndSepPunct[3]{}
\providecommand*\mciteSetBstSublistLabelBeginEnd[3]{}
\providecommand*\EndOfBibitem{}
\mciteSetBstSublistMode{f}
\mciteSetBstMaxWidthForm{subitem}{(\alph{mcitesubitemcount})}
\mciteSetBstSublistLabelBeginEnd
  {\mcitemaxwidthsubitemform\space}
  {\relax}
  {\relax}

\bibitem[Cao \latin{et~al.}(2012)Cao, Wang, Han, Ye, Zhu, Shi, Niu, Tan, Wang,
  Liu, and Feng]{Cao2012}
Cao,~T.; Wang,~G.; Han,~W.; Ye,~H.; Zhu,~C.; Shi,~J.; Niu,~Q.; Tan,~P.;
  Wang,~E.; Liu,~B. \latin{et~al.}  \emph{Nat Commun} \textbf{2012}, \emph{3},
  887--\relax
\mciteBstWouldAddEndPuncttrue
\mciteSetBstMidEndSepPunct{\mcitedefaultmidpunct}
{\mcitedefaultendpunct}{\mcitedefaultseppunct}\relax
\EndOfBibitem
\bibitem[Mak \latin{et~al.}(2012)Mak, He, Shan, and Heinz]{Mak2012}
Mak,~K.~F.; He,~K.; Shan,~J.; Heinz,~T.~F. \emph{Nat Nano} \textbf{2012},
  \emph{7}, 494--498\relax
\mciteBstWouldAddEndPuncttrue
\mciteSetBstMidEndSepPunct{\mcitedefaultmidpunct}
{\mcitedefaultendpunct}{\mcitedefaultseppunct}\relax
\EndOfBibitem
\bibitem[Zeng \latin{et~al.}(2012)Zeng, Dai, Yao, Xiao, and Cui]{Zeng2012}
Zeng,~H.; Dai,~J.; Yao,~W.; Xiao,~D.; Cui,~X. \emph{Nat Nano} \textbf{2012},
  \emph{7}, 490--493\relax
\mciteBstWouldAddEndPuncttrue
\mciteSetBstMidEndSepPunct{\mcitedefaultmidpunct}
{\mcitedefaultendpunct}{\mcitedefaultseppunct}\relax
\EndOfBibitem
\bibitem[Xu \latin{et~al.}(2014)Xu, Yao, Xiao, and Heinz]{Xu2014}
Xu,~X.; Yao,~W.; Xiao,~D.; Heinz,~T.~F. \emph{Nat Phys} \textbf{2014},
  \emph{10}, 343--350\relax
\mciteBstWouldAddEndPuncttrue
\mciteSetBstMidEndSepPunct{\mcitedefaultmidpunct}
{\mcitedefaultendpunct}{\mcitedefaultseppunct}\relax
\EndOfBibitem
\bibitem[Urbaszek and Marie(2015)Urbaszek, and Marie]{Urbaszek2015}
Urbaszek,~B.; Marie,~X. \emph{Nat Phys} \textbf{2015}, \emph{11}, 94--95\relax
\mciteBstWouldAddEndPuncttrue
\mciteSetBstMidEndSepPunct{\mcitedefaultmidpunct}
{\mcitedefaultendpunct}{\mcitedefaultseppunct}\relax
\EndOfBibitem
\bibitem[Lee \latin{et~al.}(2016)Lee, Mak, and Shan]{Lee2016}
Lee,~J.; Mak,~K.~F.; Shan,~J. \emph{Nat Nano} \textbf{2016}, \emph{11},
  421--425\relax
\mciteBstWouldAddEndPuncttrue
\mciteSetBstMidEndSepPunct{\mcitedefaultmidpunct}
{\mcitedefaultendpunct}{\mcitedefaultseppunct}\relax
\EndOfBibitem
\bibitem[Schaibley \latin{et~al.}(2016)Schaibley, Yu, Clark, Rivera, Ross,
  Seyler, Yao, and Xu]{Schaibley2016}
Schaibley,~J.~R.; Yu,~H.; Clark,~G.; Rivera,~P.; Ross,~J.~S.; Seyler,~K.~L.;
  Yao,~W.; Xu,~X. \emph{Nature Reviews Materials} \textbf{2016}, \emph{1},
  16055--16069\relax
\mciteBstWouldAddEndPuncttrue
\mciteSetBstMidEndSepPunct{\mcitedefaultmidpunct}
{\mcitedefaultendpunct}{\mcitedefaultseppunct}\relax
\EndOfBibitem
\bibitem[Glazov \latin{et~al.}(2014)Glazov, Amand, Marie, Lagarde, Bouet, and
  Urbaszek]{Glazov2014}
Glazov,~M.~M.; Amand,~T.; Marie,~X.; Lagarde,~D.; Bouet,~L.; Urbaszek,~B.
  \emph{Phys. Rev. B} \textbf{2014}, \emph{89}, 201302\relax
\mciteBstWouldAddEndPuncttrue
\mciteSetBstMidEndSepPunct{\mcitedefaultmidpunct}
{\mcitedefaultendpunct}{\mcitedefaultseppunct}\relax
\EndOfBibitem
\bibitem[Yang \latin{et~al.}(2015)Yang, Sinitsyn, Chen, Yuan, Zhang, Lou, and
  Crooker]{Yang2015}
Yang,~L.; Sinitsyn,~N.~A.; Chen,~W.; Yuan,~J.; Zhang,~J.; Lou,~J.;
  Crooker,~S.~A. \emph{Nat Phys} \textbf{2015}, \emph{11}, 830--834\relax
\mciteBstWouldAddEndPuncttrue
\mciteSetBstMidEndSepPunct{\mcitedefaultmidpunct}
{\mcitedefaultendpunct}{\mcitedefaultseppunct}\relax
\EndOfBibitem
\bibitem[Rivera \latin{et~al.}(2016)Rivera, Seyler, Yu, Schaibley, Yan,
  Mandrus, Yao, and Xu]{Rivera2016}
Rivera,~P.; Seyler,~K.~L.; Yu,~H.; Schaibley,~J.~R.; Yan,~J.; Mandrus,~D.~G.;
  Yao,~W.; Xu,~X. \emph{Science} \textbf{2016}, \emph{351}, 688--691\relax
\mciteBstWouldAddEndPuncttrue
\mciteSetBstMidEndSepPunct{\mcitedefaultmidpunct}
{\mcitedefaultendpunct}{\mcitedefaultseppunct}\relax
\EndOfBibitem
\bibitem[Ye \latin{et~al.}(2016)Ye, Xiao, Wang, Ye, Zhu, Zhao, Wang, Zhao, Yin,
  and Zhang]{Ye2016}
Ye,~Y.; Xiao,~J.; Wang,~H.; Ye,~Z.; Zhu,~H.; Zhao,~M.; Wang,~Y.; Zhao,~J.;
  Yin,~X.; Zhang,~X. \emph{Nat Nano} \textbf{2016}, \emph{11}, 598--602\relax
\mciteBstWouldAddEndPuncttrue
\mciteSetBstMidEndSepPunct{\mcitedefaultmidpunct}
{\mcitedefaultendpunct}{\mcitedefaultseppunct}\relax
\EndOfBibitem
\bibitem[Rycerz \latin{et~al.}(2007)Rycerz, Tworzydlo, and
  Beenakker]{Rycerz2007}
Rycerz,~A.; Tworzydlo,~J.; Beenakker,~C. W.~J. \emph{Nat Phys} \textbf{2007},
  \emph{3}, 172--175\relax
\mciteBstWouldAddEndPuncttrue
\mciteSetBstMidEndSepPunct{\mcitedefaultmidpunct}
{\mcitedefaultendpunct}{\mcitedefaultseppunct}\relax
\EndOfBibitem
\bibitem[Xiao \latin{et~al.}(2012)Xiao, Liu, Feng, Xu, and Yao]{Xiao2012}
Xiao,~D.; Liu,~G.-B.; Feng,~W.; Xu,~X.; Yao,~W. \emph{Phys. Rev. Lett.}
  \textbf{2012}, \emph{108}, 196802\relax
\mciteBstWouldAddEndPuncttrue
\mciteSetBstMidEndSepPunct{\mcitedefaultmidpunct}
{\mcitedefaultendpunct}{\mcitedefaultseppunct}\relax
\EndOfBibitem
\bibitem[Marie and Urbaszek(2015)Marie, and Urbaszek]{Marie2015}
Marie,~X.; Urbaszek,~B. \emph{Nat Mater} \textbf{2015}, \emph{14},
  860--861\relax
\mciteBstWouldAddEndPuncttrue
\mciteSetBstMidEndSepPunct{\mcitedefaultmidpunct}
{\mcitedefaultendpunct}{\mcitedefaultseppunct}\relax
\EndOfBibitem
\bibitem[Poellmann \latin{et~al.}(2015)Poellmann, Steinleitner, Leierseder,
  Nagler, Plechinger, Porer, Bratschitsch, Schuller, Korn, and
  Huber]{Poellmann2015}
Poellmann,~C.; Steinleitner,~P.; Leierseder,~U.; Nagler,~P.; Plechinger,~G.;
  Porer,~M.; Bratschitsch,~R.; Schuller,~C.; Korn,~T.; Huber,~R. \emph{Nat
  Mater} \textbf{2015}, \emph{14}, 889--893\relax
\mciteBstWouldAddEndPuncttrue
\mciteSetBstMidEndSepPunct{\mcitedefaultmidpunct}
{\mcitedefaultendpunct}{\mcitedefaultseppunct}\relax
\EndOfBibitem
\bibitem[Nie \latin{et~al.}(2014)Nie, Long, Sun, Huang, Zhang, Xiong, Hewak,
  Shen, Prezhdo, and Loh]{Nie2014}
Nie,~Z.; Long,~R.; Sun,~L.; Huang,~C.-C.; Zhang,~J.; Xiong,~Q.; Hewak,~D.~W.;
  Shen,~Z.; Prezhdo,~O.~V.; Loh,~Z.-H. \emph{ACS Nano} \textbf{2014}, \emph{8},
  10931--10940\relax
\mciteBstWouldAddEndPuncttrue
\mciteSetBstMidEndSepPunct{\mcitedefaultmidpunct}
{\mcitedefaultendpunct}{\mcitedefaultseppunct}\relax
\EndOfBibitem
\bibitem[Zhukov \latin{et~al.}(2007)Zhukov, Yakovlev, Bayer, Glazov, Ivchenko,
  Karczewski, Wojtowicz, and Kossut]{Zhukov2007}
Zhukov,~E.~A.; Yakovlev,~D.~R.; Bayer,~M.; Glazov,~M.~M.; Ivchenko,~E.~L.;
  Karczewski,~G.; Wojtowicz,~T.; Kossut,~J. \emph{Phys. Rev. B} \textbf{2007},
  \emph{76}, 205310\relax
\mciteBstWouldAddEndPuncttrue
\mciteSetBstMidEndSepPunct{\mcitedefaultmidpunct}
{\mcitedefaultendpunct}{\mcitedefaultseppunct}\relax
\EndOfBibitem
\bibitem[Zhu \latin{et~al.}(2014)Zhu, Zhang, Glazov, Urbaszek, Amand, Ji, Liu,
  and Marie]{Zhu2014}
Zhu,~C.~R.; Zhang,~K.; Glazov,~M.; Urbaszek,~B.; Amand,~T.; Ji,~Z.~W.;
  Liu,~B.~L.; Marie,~X. \emph{Phys. Rev. B} \textbf{2014}, \emph{90},
  161302\relax
\mciteBstWouldAddEndPuncttrue
\mciteSetBstMidEndSepPunct{\mcitedefaultmidpunct}
{\mcitedefaultendpunct}{\mcitedefaultseppunct}\relax
\EndOfBibitem
\bibitem[Dal~Conte \latin{et~al.}(2015)Dal~Conte, Bottegoni, Pogna, De~Fazio,
  Ambrogio, Bargigia, D'Andrea, Lombardo, Bruna, Ciccacci, Ferrari, Cerullo,
  and Finazzi]{DalConte2015}
Dal~Conte,~S.; Bottegoni,~F.; Pogna,~E. A.~A.; De~Fazio,~D.; Ambrogio,~S.;
  Bargigia,~I.; D'Andrea,~C.; Lombardo,~A.; Bruna,~M.; Ciccacci,~F.
  \latin{et~al.}  \emph{Phys. Rev. B} \textbf{2015}, \emph{92}, 235425\relax
\mciteBstWouldAddEndPuncttrue
\mciteSetBstMidEndSepPunct{\mcitedefaultmidpunct}
{\mcitedefaultendpunct}{\mcitedefaultseppunct}\relax
\EndOfBibitem
\bibitem[Winnerl \latin{et~al.}(2011)Winnerl, Orlita, Plochocka, Kossacki,
  Potemski, Winzer, Malic, Knorr, Sprinkle, Berger, de~Heer, Schneider, and
  Helm]{Winnerl2011}
Winnerl,~S.; Orlita,~M.; Plochocka,~P.; Kossacki,~P.; Potemski,~M.; Winzer,~T.;
  Malic,~E.; Knorr,~A.; Sprinkle,~M.; Berger,~C. \latin{et~al.}  \emph{Phys.
  Rev. Lett.} \textbf{2011}, \emph{107}, 237401\relax
\mciteBstWouldAddEndPuncttrue
\mciteSetBstMidEndSepPunct{\mcitedefaultmidpunct}
{\mcitedefaultendpunct}{\mcitedefaultseppunct}\relax
\EndOfBibitem
\bibitem[Mai \latin{et~al.}(2014)Mai, Barrette, Yu, Semenov, Kim, Cao, and
  Gundogdu]{Mai2014a}
Mai,~C.; Barrette,~A.; Yu,~Y.; Semenov,~Y.~G.; Kim,~K.~W.; Cao,~L.;
  Gundogdu,~K. \emph{Nano Letters} \textbf{2014}, \emph{14}, 202--206\relax
\mciteBstWouldAddEndPuncttrue
\mciteSetBstMidEndSepPunct{\mcitedefaultmidpunct}
{\mcitedefaultendpunct}{\mcitedefaultseppunct}\relax
\EndOfBibitem
\bibitem[Yu and Wu(2014)Yu, and Wu]{yu_valley_2014}
Yu,~T.; Wu,~M.~W. \emph{Phys. Rev. B} \textbf{2014}, \emph{89}, 205303\relax
\mciteBstWouldAddEndPuncttrue
\mciteSetBstMidEndSepPunct{\mcitedefaultmidpunct}
{\mcitedefaultendpunct}{\mcitedefaultseppunct}\relax
\EndOfBibitem
\bibitem[Glazov \latin{et~al.}(2015)Glazov, Ivchenko, Wang, Amand, Marie,
  Urbaszek, and Liu]{Glazov2015}
Glazov,~M.~M.; Ivchenko,~E.~L.; Wang,~G.; Amand,~T.; Marie,~X.; Urbaszek,~B.;
  Liu,~B.~L. \emph{physica status solidi (b)} \textbf{2015}, \emph{252},
  2349--2362\relax
\mciteBstWouldAddEndPuncttrue
\mciteSetBstMidEndSepPunct{\mcitedefaultmidpunct}
{\mcitedefaultendpunct}{\mcitedefaultseppunct}\relax
\EndOfBibitem
\bibitem[Sie \latin{et~al.}(2015)Sie, McIver, Lee, Fu, Kong, and
  Gedik]{Sie2015}
Sie,~E.~J.; McIver,~J.~W.; Lee,~Y.-H.; Fu,~L.; Kong,~J.; Gedik,~N. \emph{Nat
  Mater} \textbf{2015}, \emph{14}, 290--294\relax
\mciteBstWouldAddEndPuncttrue
\mciteSetBstMidEndSepPunct{\mcitedefaultmidpunct}
{\mcitedefaultendpunct}{\mcitedefaultseppunct}\relax
\EndOfBibitem
\bibitem[Hao \latin{et~al.}(2016)Hao, Moody, Wu, Dass, Xu, Chen, Sun, Li, Li,
  MacDonald, and Li]{Hao2016}
Hao,~K.; Moody,~G.; Wu,~F.; Dass,~C.~K.; Xu,~L.; Chen,~C.-H.; Sun,~L.;
  Li,~M.-Y.; Li,~L.-J.; MacDonald,~A.~H. \latin{et~al.}  \emph{Nat Phys}
  \textbf{2016}, \emph{12}, 677--682\relax
\mciteBstWouldAddEndPuncttrue
\mciteSetBstMidEndSepPunct{\mcitedefaultmidpunct}
{\mcitedefaultendpunct}{\mcitedefaultseppunct}\relax
\EndOfBibitem
\bibitem[Maialle \latin{et~al.}(1993)Maialle, de~Andrada~e Silva, and
  Sham]{Maialle1993}
Maialle,~M.~Z.; de~Andrada~e Silva,~E.~A.; Sham,~L.~J. \emph{Phys. Rev. B}
  \textbf{1993}, \emph{47}, 15776--15788\relax
\mciteBstWouldAddEndPuncttrue
\mciteSetBstMidEndSepPunct{\mcitedefaultmidpunct}
{\mcitedefaultendpunct}{\mcitedefaultseppunct}\relax
\EndOfBibitem
\bibitem[Molina-S\'{a}nchez \latin{et~al.}(2015)Molina-S\'{a}nchez, Hummer, and
  Wirtz]{Molina-Sanchez2015}
Molina-S\'{a}nchez,~A.; Hummer,~K.; Wirtz,~L. \emph{Surface Science Reports}
  \textbf{2015}, \emph{70}, 554 -- 586\relax
\mciteBstWouldAddEndPuncttrue
\mciteSetBstMidEndSepPunct{\mcitedefaultmidpunct}
{\mcitedefaultendpunct}{\mcitedefaultseppunct}\relax
\EndOfBibitem
\bibitem[Molina-S\'anchez \latin{et~al.}(2016)Molina-S\'anchez, Palummo,
  Marini, and Wirtz]{Molina-Sanchez2016}
Molina-S\'anchez,~A.; Palummo,~M.; Marini,~A.; Wirtz,~L. \emph{Phys. Rev. B}
  \textbf{2016}, \emph{93}, 155435\relax
\mciteBstWouldAddEndPuncttrue
\mciteSetBstMidEndSepPunct{\mcitedefaultmidpunct}
{\mcitedefaultendpunct}{\mcitedefaultseppunct}\relax
\EndOfBibitem
\bibitem[Cudazzo \latin{et~al.}(2011)Cudazzo, Tokatly, and Rubio]{Cudazzo2011}
Cudazzo,~P.; Tokatly,~I.~V.; Rubio,~A. \emph{Phys. Rev. B} \textbf{2011},
  \emph{84}, 085406\relax
\mciteBstWouldAddEndPuncttrue
\mciteSetBstMidEndSepPunct{\mcitedefaultmidpunct}
{\mcitedefaultendpunct}{\mcitedefaultseppunct}\relax
\EndOfBibitem
\bibitem[Latini \latin{et~al.}(2015)Latini, Olsen, and Thygesen]{Latini2015}
Latini,~S.; Olsen,~T.; Thygesen,~K.~S. \emph{Phys. Rev. B} \textbf{2015},
  \emph{92}, 245123\relax
\mciteBstWouldAddEndPuncttrue
\mciteSetBstMidEndSepPunct{\mcitedefaultmidpunct}
{\mcitedefaultendpunct}{\mcitedefaultseppunct}\relax
\EndOfBibitem
\bibitem[Onida \latin{et~al.}(2002)Onida, Reining, and Rubio]{Onida2002}
Onida,~G.; Reining,~L.; Rubio,~A. \emph{Rev. Mod. Phys.} \textbf{2002},
  \emph{74}, 601--659\relax
\mciteBstWouldAddEndPuncttrue
\mciteSetBstMidEndSepPunct{\mcitedefaultmidpunct}
{\mcitedefaultendpunct}{\mcitedefaultseppunct}\relax
\EndOfBibitem
\bibitem[Wu \latin{et~al.}(2013)Wu, Ross, Liu, Aivazian, Jones, Fei, Zhu, Xiao,
  Yao, Cobden, and Xu]{Wu2013}
Wu,~S.; Ross,~J.~S.; Liu,~G.-B.; Aivazian,~G.; Jones,~A.; Fei,~Z.; Zhu,~W.;
  Xiao,~D.; Yao,~W.; Cobden,~D. \latin{et~al.}  \emph{Nat Phys} \textbf{2013},
  \emph{9}, 149--153\relax
\mciteBstWouldAddEndPuncttrue
\mciteSetBstMidEndSepPunct{\mcitedefaultmidpunct}
{\mcitedefaultendpunct}{\mcitedefaultseppunct}\relax
\EndOfBibitem
\bibitem[Xie and Cui(2016)Xie, and Cui]{Xie2016}
Xie,~L.; Cui,~X. \emph{Proceedings of the National Academy of Sciences}
  \textbf{2016}, \emph{113}, 3746--3750\relax
\mciteBstWouldAddEndPuncttrue
\mciteSetBstMidEndSepPunct{\mcitedefaultmidpunct}
{\mcitedefaultendpunct}{\mcitedefaultseppunct}\relax
\EndOfBibitem
\bibitem[Marini(2013)]{Marini2013}
Marini,~A. \emph{Journal of Physics: Conference Series} \textbf{2013},
  \emph{427}, 012003\relax
\mciteBstWouldAddEndPuncttrue
\mciteSetBstMidEndSepPunct{\mcitedefaultmidpunct}
{\mcitedefaultendpunct}{\mcitedefaultseppunct}\relax
\EndOfBibitem
\bibitem[Sangalli and Marini(2015)Sangalli, and Marini]{Sangalli2015}
Sangalli,~D.; Marini,~A. \emph{EPL (Europhysics Letters)} \textbf{2015},
  \emph{110}, 47004\relax
\mciteBstWouldAddEndPuncttrue
\mciteSetBstMidEndSepPunct{\mcitedefaultmidpunct}
{\mcitedefaultendpunct}{\mcitedefaultseppunct}\relax
\EndOfBibitem
\bibitem[Sangalli \latin{et~al.}(2016)Sangalli, Dal~Conte, Manzoni, Cerullo,
  and Marini]{Sangalli2016}
Sangalli,~D.; Dal~Conte,~S.; Manzoni,~C.; Cerullo,~G.; Marini,~A. \emph{Phys.
  Rev. B} \textbf{2016}, \emph{93}, 195205\relax
\mciteBstWouldAddEndPuncttrue
\mciteSetBstMidEndSepPunct{\mcitedefaultmidpunct}
{\mcitedefaultendpunct}{\mcitedefaultseppunct}\relax
\EndOfBibitem
\bibitem[de~Melo and Marini(2016)de~Melo, and Marini]{Melo2016}
de~Melo,~P. M. M.~C.; Marini,~A. \emph{Phys. Rev. B} \textbf{2016}, \emph{93},
  155102\relax
\mciteBstWouldAddEndPuncttrue
\mciteSetBstMidEndSepPunct{\mcitedefaultmidpunct}
{\mcitedefaultendpunct}{\mcitedefaultseppunct}\relax
\EndOfBibitem
\bibitem[Attaccalite \latin{et~al.}(2011)Attaccalite, Gr\"uning, and
  Marini]{Attaccalite2011}
Attaccalite,~C.; Gr\"uning,~M.; Marini,~A. \emph{Phys. Rev. B} \textbf{2011},
  \emph{84}, 245110\relax
\mciteBstWouldAddEndPuncttrue
\mciteSetBstMidEndSepPunct{\mcitedefaultmidpunct}
{\mcitedefaultendpunct}{\mcitedefaultseppunct}\relax
\EndOfBibitem
\bibitem[Pogna \latin{et~al.}(2016)Pogna, Marsili, Fazio, Conte, Manzoni,
  Sangalli, Yoon, Lombardo, Ferrari, Marini, Cerullo, and Prezzi]{Pogna2016}
Pogna,~E. A.~A.; Marsili,~M.; Fazio,~D.~D.; Conte,~S.~D.; Manzoni,~C.;
  Sangalli,~D.; Yoon,~D.; Lombardo,~A.; Ferrari,~A.~C.; Marini,~A.
  \latin{et~al.}  \emph{ACS Nano} \textbf{2016}, \emph{10}, 1182--1188\relax
\mciteBstWouldAddEndPuncttrue
\mciteSetBstMidEndSepPunct{\mcitedefaultmidpunct}
{\mcitedefaultendpunct}{\mcitedefaultseppunct}\relax
\EndOfBibitem
\bibitem[Perfetto \latin{et~al.}(2016)Perfetto, Sangalli, Marini, and
  Stefanucci]{Perfetto2016}
Perfetto,~E.; Sangalli,~D.; Marini,~A.; Stefanucci,~G. \emph{Phys. Rev. B}
  \textbf{2016}, \emph{94}, 245303\relax
\mciteBstWouldAddEndPuncttrue
\mciteSetBstMidEndSepPunct{\mcitedefaultmidpunct}
{\mcitedefaultendpunct}{\mcitedefaultseppunct}\relax
\EndOfBibitem
\bibitem[Gr\"uning and Attaccalite(2014)Gr\"uning, and
  Attaccalite]{PhysRevB.89.081102}
Gr\"uning,~M.; Attaccalite,~C. \emph{Phys. Rev. B} \textbf{2014}, \emph{89},
  081102\relax
\mciteBstWouldAddEndPuncttrue
\mciteSetBstMidEndSepPunct{\mcitedefaultmidpunct}
{\mcitedefaultendpunct}{\mcitedefaultseppunct}\relax
\EndOfBibitem
\bibitem[Attaccalite and Gr\"uning(2013)Attaccalite, and
  Gr\"uning]{PhysRevB.88.235113}
Attaccalite,~C.; Gr\"uning,~M. \emph{Phys. Rev. B} \textbf{2013}, \emph{88},
  235113\relax
\mciteBstWouldAddEndPuncttrue
\mciteSetBstMidEndSepPunct{\mcitedefaultmidpunct}
{\mcitedefaultendpunct}{\mcitedefaultseppunct}\relax
\EndOfBibitem
\bibitem[Attaccalite \latin{et~al.}(2015)Attaccalite, Nguer, Cannuccia, and
  Gr\"{u}ning]{C5CP00601E}
Attaccalite,~C.; Nguer,~A.; Cannuccia,~E.; Gr\"{u}ning,~M. \emph{Phys. Chem.
  Chem. Phys.} \textbf{2015}, \emph{17}, 9533--9540\relax
\mciteBstWouldAddEndPuncttrue
\mciteSetBstMidEndSepPunct{\mcitedefaultmidpunct}
{\mcitedefaultendpunct}{\mcitedefaultseppunct}\relax
\EndOfBibitem
\bibitem[Marini(2008)]{Marini2008}
Marini,~A. \emph{Phys. Rev. Lett.} \textbf{2008}, \emph{101}, 106405\relax
\mciteBstWouldAddEndPuncttrue
\mciteSetBstMidEndSepPunct{\mcitedefaultmidpunct}
{\mcitedefaultendpunct}{\mcitedefaultseppunct}\relax
\EndOfBibitem
\bibitem[Sangalli and Marini(2015)Sangalli, and Marini]{Sangalli2015a}
Sangalli,~D.; Marini,~A. \emph{Journal of Physics: Conference Series}
  \textbf{2015}, \emph{609}, 012006\relax
\mciteBstWouldAddEndPuncttrue
\mciteSetBstMidEndSepPunct{\mcitedefaultmidpunct}
{\mcitedefaultendpunct}{\mcitedefaultseppunct}\relax
\EndOfBibitem
\bibitem[Perfetto \latin{et~al.}(2015)Perfetto, Sangalli, Marini, and
  Stefanucci]{Perfetto2015}
Perfetto,~E.; Sangalli,~D.; Marini,~A.; Stefanucci,~G. \emph{Phys. Rev. B}
  \textbf{2015}, \emph{92}, 205304\relax
\mciteBstWouldAddEndPuncttrue
\mciteSetBstMidEndSepPunct{\mcitedefaultmidpunct}
{\mcitedefaultendpunct}{\mcitedefaultseppunct}\relax
\EndOfBibitem
\bibitem[Sangalli \latin{et~al.}(2012)Sangalli, Marini, and
  Debernardi]{Sangalli2012}
Sangalli,~D.; Marini,~A.; Debernardi,~A. \emph{Phys. Rev. B} \textbf{2012},
  \emph{86}, 125139\relax
\mciteBstWouldAddEndPuncttrue
\mciteSetBstMidEndSepPunct{\mcitedefaultmidpunct}
{\mcitedefaultendpunct}{\mcitedefaultseppunct}\relax
\EndOfBibitem
\bibitem[Zhang \latin{et~al.}(2015)Zhang, You, Zhao, and Heinz]{Zhang2015}
Zhang,~X.-X.; You,~Y.; Zhao,~S. Y.~F.; Heinz,~T.~F. \emph{Phys. Rev. Lett.}
  \textbf{2015}, \emph{115}, 257403\relax
\mciteBstWouldAddEndPuncttrue
\mciteSetBstMidEndSepPunct{\mcitedefaultmidpunct}
{\mcitedefaultendpunct}{\mcitedefaultseppunct}\relax
\EndOfBibitem
\bibitem[Robert \latin{et~al.}(2016)Robert, Lagarde, Cadiz, Wang, Lassagne,
  Amand, Balocchi, Renucci, Tongay, Urbaszek, and Marie]{Robert2016}
Robert,~C.; Lagarde,~D.; Cadiz,~F.; Wang,~G.; Lassagne,~B.; Amand,~T.;
  Balocchi,~A.; Renucci,~P.; Tongay,~S.; Urbaszek,~B. \latin{et~al.}
  \emph{Phys. Rev. B} \textbf{2016}, \emph{93}, 205423\relax
\mciteBstWouldAddEndPuncttrue
\mciteSetBstMidEndSepPunct{\mcitedefaultmidpunct}
{\mcitedefaultendpunct}{\mcitedefaultseppunct}\relax
\EndOfBibitem
\bibitem[Palummo \latin{et~al.}(2015)Palummo, Bernardi, and
  Grossman]{Palummo2015}
Palummo,~M.; Bernardi,~M.; Grossman,~J.~C. \emph{Nano Letters} \textbf{2015},
  \emph{15}, 2794--2800\relax
\mciteBstWouldAddEndPuncttrue
\mciteSetBstMidEndSepPunct{\mcitedefaultmidpunct}
{\mcitedefaultendpunct}{\mcitedefaultseppunct}\relax
\EndOfBibitem
\bibitem[Bernardi \latin{et~al.}(2014)Bernardi, Vigil-Fowler, Lischner, Neaton,
  and Louie]{Bernardi2014}
Bernardi,~M.; Vigil-Fowler,~D.; Lischner,~J.; Neaton,~J.~B.; Louie,~S.~G.
  \emph{Phys. Rev. Lett.} \textbf{2014}, \emph{112}, 257402\relax
\mciteBstWouldAddEndPuncttrue
\mciteSetBstMidEndSepPunct{\mcitedefaultmidpunct}
{\mcitedefaultendpunct}{\mcitedefaultseppunct}\relax
\EndOfBibitem
\bibitem[spi()]{spinflip}
In the strict sense, spin is only a good quantum number exactly at $K^+$ and
  $K^-$. In the region surrounding $K^+$ and $K^-$, the electron/hole states
  are not exact eigenspinors of $\hat{s}_z$, making possible such a
  \textit{spin-flip} transition.\relax
\mciteBstWouldAddEndPunctfalse
\mciteSetBstMidEndSepPunct{\mcitedefaultmidpunct}
{}{\mcitedefaultseppunct}\relax
\EndOfBibitem
\bibitem[Kioseoglou \latin{et~al.}(2016)Kioseoglou, Hanbicki, Currie, Friedman,
  and Jonker]{Kioseoglou2016}
Kioseoglou,~G.; Hanbicki,~A.~T.; Currie,~M.; Friedman,~A.~L.; Jonker,~B.~T.
  \emph{Scientific Reports} \textbf{2016}, \emph{6}, 25041--\relax
\mciteBstWouldAddEndPuncttrue
\mciteSetBstMidEndSepPunct{\mcitedefaultmidpunct}
{\mcitedefaultendpunct}{\mcitedefaultseppunct}\relax
\EndOfBibitem
\bibitem[Varrette \latin{et~al.}(2014)Varrette, Bouvry, Cartiaux, and
  Georgatos]{VBCG_HPCS14}
Varrette,~S.; Bouvry,~P.; Cartiaux,~H.; Georgatos,~F. Management of an Academic
  HPC Cluster: The UL Experience. Proc. of the 2014 Intl. Conf. on High
  Performance Computing \& Simulation (HPCS 2014). Bologna, Italy, 2014; pp
  959--967\relax
\mciteBstWouldAddEndPuncttrue
\mciteSetBstMidEndSepPunct{\mcitedefaultmidpunct}
{\mcitedefaultendpunct}{\mcitedefaultseppunct}\relax
\EndOfBibitem
\end{mcitethebibliography}
\providecommand{\latin}[1]{#1}
\providecommand*\mcitethebibliography{\thebibliography}
\csname @ifundefined\endcsname{endmcitethebibliography}
  {\let\endmcitethebibliography\endthebibliography}{}

\end{document}